\documentclass[11pt,a4paper]{article}
\pdfoutput=1
\usepackage{jheppub}
\usepackage{amsmath}%
\usepackage{amsfonts}%
\usepackage{amssymb}%
\usepackage{graphicx}
\usepackage{hyperref}
\usepackage{caption}
\usepackage{subcaption}

\begin{document}

\title{Mutual information on the fuzzy sphere}
\author{Philippe Sabella-Garnier}
\affiliation{Department of Physics and Astronomy, University of British Columbia, \\ 6224 Agricultural Road,
Vancouver, Canada}
\emailAdd{psabella@phas.ubc.ca}
\abstract{We numerically calculate entanglement entropy and mutual information for a massive free scalar field on commutative (ordinary) and noncommutative (fuzzy) spheres. We regularize the theory on the commutative geometry by discretizing the polar coordinate, whereas the theory on the noncommutative geometry naturally posseses a finite and adjustable number of degrees of freedom. Our results show that the UV-divergent part of the entanglement entropy on a fuzzy sphere does not follow an area law, while the entanglement entropy on a commutative sphere does. Nonetheless, we find that mutual information (which is UV-finite) is the same in both theories. This suggests that nonlocality at short distances does not affect quantum correlations over large distances in a free field theory.}
\arxivnumber{1409.7069}

\maketitle

\section{Introduction}
Entanglement entropy has attracted a lot of interest in the string theory community because of the connection it makes between a well-defined classical geometric quantity and a purely quantum phenomenon through the Ryu-Takayanagi proposal \cite{Ryu:2006bv}:
\begin{equation}
S_A=\frac{\text{Area}(\gamma_A)}{4G_N} ~, 
\end{equation} 
where $S_A$ is the entanglement entropy of a region $A$ in a CFT and $\gamma_A$ is an extremal surface in the bulk of the dual space-time that has the same boundary as $A$. In addition to the obvious application of learning something new about strongly-coupled field theories, this proposal has been used to understand gravity as a consequence of information-theoretic considerations \cite{Faulkner:2013, Swingle:2014}. In the same spirit (but from a field theory point of view), we would like to use entanglement entropy to better understand how the degrees of freedom of field theories on noncommutative geometries are distributed.

Locality makes the leading-order UV divergence of the entanglement entropy between two regions scale like the area of the boundary between them \cite{Srednicki:1993im, Eisert:2008ur}, so we can use entanglement entropy to probe the degree of non-locality of a quantum field theory at short distances. Conversely, mutual information is calculated from the finite parts of entanglement entropy. It provides a bound on the range of correlations \cite{Groisman:2005} and can therefore be seen as a measure of non-locality in the IR. 

Holographic calculations of entanglement entropy on noncommutative geometries have shown a departure from the area law \cite{Fischler:2013gsa, Karczmarek:2013}. Furthermore it was also shown in \cite{Karczmarek:2013} that mutual information for strongly-coupled theories behaves differently in commutative and non-commutative theory due to UV/IR mixing. In \cite{Karczmarek:2014}, it was shown that a departure from an area law behaviour can also be seen on the fuzzy sphere through a field theory calculation.\footnote{For previous calculations of entanglement entropy on the fuzzy sphere, see \cite{Dou:2006ni,Dou:2009cw}} The effect was shown to persist even in the so-called ``commutative'' limit, another sign of an UV/IR connection in entanglement entropy.

In this paper, we examine the properties of mutual information for a massive free scalar field on both a commutative and a noncommutative sphere in the hope that the comparison will teach us something about the distribution of degrees of freedom and the IR structure of the theory. The commutative theory is regularized by discretizing the polar angle $\theta$. We find that mutual information between two regions separated by an annulus of variable width is the same for the commutative and noncommutative theories both when the theory is conformal and when it is not.

The remainder of this paper is organized as follows. In section 2, we introduce our procedure to regularize the commutative sphere and briefly recall the construction of field theory on the fuzzy sphere. In section 3, we review the leading-order behaviour of entanglement entropy before examining mutual information on both theories. We conclude in section 4 by discussing some of the implications of our results and possible future work.

\section{Actions and numerical set-up}
Consider a quantum-mechanical system living in a Hilbert space $\mathcal{H}=\mathcal{H}_A \otimes \mathcal{H}_{\bar{A}}$, with ground state $|\Psi \rangle$. An observer living in $A$, for whom $\bar{A}$ is inaccessible, sees the ground state as a density matrix defined as
\begin{equation}
\rho_A=\text{Tr}_{\bar{A}} \left(| \Psi \rangle \langle \Psi |\right)~.
\end{equation}
The entanglement entropy (see, e.g. \cite{Nishioka:2009un}) of $A$ is then the von Neumann entropy of $\rho_A$:
\begin{equation}
S_A=-\text{Tr}_A \left(\rho_A \log \rho_A \right)~.
\end{equation}
Given a Hamiltonian of the form
\begin{equation}
H=\frac{1}{2}\sum_{i,j=1}^{N} \left(\delta_{ij}p_i p_j + x_i K_{ij} x_j\right) ~,
\label{Hform}
\end{equation}
with $K_{ij}$ a positive-definite matrix and $[x_i,p_j]=i\delta_{ij}$, it is a straightforward exercise to calculate the ground-state entanglement entropy associated with the ``  region'' $i \leq I$ (\cite{Srednicki:1993im},\cite{Casini:2009}). Simply define
\begin{equation}
[X_I]_{ij}=\frac{1}{2}[K^{-1/2}]_{ij} \; \; \; [P_I]_{ij}=\frac{1}{2}[K^{1/2}]_{ij} \; \; \; i,j \leq I ~,
\end{equation}
and the entanglement entropy is 
\begin{equation}
S_I=\text{Tr}\left\{ (X_I\cdot P_I + \frac{1}{2} \mathbb{I})\log(X_I\cdot P_I + \frac{1}{2} \mathbb{I}) - (X_I\cdot P_I - \frac{1}{2} \mathbb{I})\log(X_I\cdot P_I - \frac{1}{2} \mathbb{I}) \right\} ~.
\label{eenum}
\end{equation}
\\

The mutual information between two disjoint regions $A$ and $B$ is defined as
\begin{equation}
I(A,B)=S_A+S_B - S_{A\cup B} ~,
\end{equation}
and can therefore be calculated using (\ref{eenum}) as well.

\subsection{Regularized Sphere}
We wish to calculate the entanglement entropy between a polar cap and its complement for the ground state of a real scalar field on a sphere of radius 1. This is a divergent quantity, therefore we must start by regularizing the field theory. The most natural regularization scheme on a spherical geometry is to expand functions in spherical harmonics and cut off the expansion at some highest mode $N$. However, this is not the most useful procedure in this situation for two reasons: the Hamiltonian expressed in terms of spherical harmonics modes is diagonal (which is usually a desirable feature but leads to zero entanglement entropy) and there is no simple way to associate contiguous regions to ranges of modes.

Given that the regions that interest us are polar caps, the next most natural regularization scheme (in the spirit of the one presented in \cite{Srednicki:1993im}) is simply to cut up the continuous polar $\theta$ variable into an evenly-spaced mesh and expand the azimuthal coordinate $\phi$ in spherical harmonics. 
The field theory Hamiltonian is

\begin{equation}
H=\frac{1}{2}\int{d\Omega \left(\Pi^2 + |\nabla \Phi|^2 + \mu^2 \Phi^2 \right) } ~,
\end{equation}
with the usual canonical commutation relation $[\Phi(\Omega),\Pi(\Omega')]=i\delta(\Omega-\Omega')$. We can write $H$ as:
\begin{equation}
H=\frac{1}{2}\int{d\Omega\left(\Pi^2+ \left(\frac{\partial \Phi}{\partial \theta}\right)^2+ \frac{1}{\sin^2\theta} \left(\frac{\partial \Phi}{\partial \phi}\right)^2   +\mu^2 \Phi^2    \right)} ~.
\end{equation}

Define:\footnote{A similar scheme was presented in \cite{Herzog:2014}.}
\begin{align}
\Phi(\theta,\phi)=\frac{1}{\sqrt{\pi \sin\theta}}\left[\frac{b_0}{\sqrt{2}}+ \sum_{m=1}^{\infty} \left( a_m \sin m\phi + b_m \cos m\phi\right)\right] ~, \\ 
\Pi(\theta,\phi)=\frac{1}{\sqrt{\pi \sin\theta}} \left[\frac{d_0}{\sqrt{2}}+ \sum_{m=1}^{\infty} \left( c_m \sin m\phi + d_m \cos m\phi\right) \right] ~.
\end{align}
One can check that the Fourier coefficients are ($m>0$):

\begin{eqnarray}
&a_m=\sqrt{\frac{\sin\theta}{\pi}}\int_0^{2\pi} d\phi \Phi(\theta,\phi) \sin m \phi ~, 
\quad b_m=\sqrt{\frac{\sin\theta}{\pi}}\int_0^{2\pi} d\phi \Phi(\theta,\phi) \cos m \phi ~, \nonumber\\
&c_m=\sqrt{\frac{\sin\theta}{\pi}}\int_0^{2\pi} d\phi \Pi(\theta,\phi) \sin m \phi ~, \quad
d_m=\sqrt{\frac{\sin\theta}{\pi}}\int_0^{2\pi} d\phi \Pi(\theta,\phi) \cos m \phi ~, \nonumber \\
&b_0= \sqrt{\frac{\sin\theta}{2\pi}}\int_0^{2\pi} d\phi \Phi(\theta,\phi) ~, \quad
d_0=\sqrt{\frac{\sin\theta}{2\pi}}\int_0^{2\pi} d\phi \Pi(\theta,\phi)~,
\end{eqnarray}

and that the non-vanishing commutation relations are:
\begin{equation}
[a_m(\theta),c_{m'}(\theta')]=[b_m(\theta),d_{m'}(\theta')]=i\delta(\theta-\theta')\delta_{mm'} ~.
\end{equation}
The first term of the Hamiltonian is (taking $a_0=c_0=0$ for simplicity of notation):
\begin{equation}
\frac{1}{2}  \int d\theta \sum_{m=0}^{\infty}( c_m ^2 + d_m^2 ) ~.
\end{equation}
The second term is:
\begin{equation}
\frac{1}{2} \int d\theta \sin \theta \sum_{m=0}^{\infty} \left[ \left( \frac{\partial}{\partial \theta} \frac{a_m}{\sqrt{\sin \theta}} \right)^2+ \left( \frac{\partial}{\partial \theta} \frac{b_m}{\sqrt{\sin \theta}} \right)^2 \right]~,
\end{equation}
and the third and fourth are:
\begin{equation}
\frac{1}{2} \sum_{m=0}^{\infty} \int d\theta \left(\frac{m^2}{\sin^2\theta}+\mu^2\right) \left( a_m^2+b_m^2\right) 
\end{equation}
We relabel our terms so that 
\begin{eqnarray}
\Phi_m=b_m, \; \; \Pi_m=d_m  \; \; \; \; (m \geq 0) \nonumber \\
 \Phi_{m}=a_{-m}, \; \; \Pi_{m}=c_{-m} \; \; \; \; (m < 0) ~.
\end{eqnarray}
Now, we discretize the polar coordinate: 
\begin{equation}
\theta \rightarrow \theta_n=n\frac{\pi}{N} \; \; \; \; n=1\dots N-1 ~.
\end{equation}
Since our continuous coordinates are now approximated by a mesh, we must replace the integrals above with Riemann sums. For the first and third terms, we use the trapezoidal rule: the integral is approximated as the average of the left and right Riemann sums, with an error of $\mathcal{O}(1/N^2)$ (as opposed to $\mathcal{O}(1/N)$ for just a left or a right sum). For the second term, we pick a middle Riemann sum, evaluating the summands at the half-point of each interval. This also has an error of $\mathcal{O}(1/N^2)$. The above terms become:

\begin{eqnarray}
&&\frac{1}{2} \sum_{m=-\infty}^{\infty} \frac{\pi}{N} \left(\frac{1}{2}\Pi_m(\theta_{1})^2+\sum_{n=2}^{N-2} \Pi_m(\theta_n) ^2 +\frac{1}{2}\Pi_m(\theta_{N-1})^2 + \mathcal{O}(1/N^2) \right)  ~, \ \\
&&\frac{1}{2} \sum_{m=-\infty}^{\infty} \left(\sum_{n=1}^{N-2} \frac{\pi}{N} \sin \theta_{n+\frac{1}{2}} \left[ \left( \frac{\partial}{\partial \theta} \frac{\Phi_m(\theta)}{\sqrt{\sin \theta}} \right)^2\right]_{\theta=\theta_{n+\frac{1}{2}}} + \mathcal{O}(1/N^2)  \right) ~,
\end{eqnarray}
\begin{equation}
\begin{split}
\frac{1}{2} \sum_{m=-\infty}^{\infty}\frac{\pi}{N} \left(\frac{1}{2}\left(\frac{m^2}{\sin^2\theta_1}+\mu^2\right)\Phi_m(\theta_1)^2  + \sum_{n=2}^{N-2}  \left(\frac{m^2}{\sin^2\theta_n}+\mu^2\right)  \Phi_m(\theta_n)^2 + \right. \\ \left. \frac{1}{2}\left(\frac{m^2}{\sin^2\theta_{N-1}}+\mu^2\right)\Phi_m(\theta_{N-1})^2+\mathcal{O}(1/N^2)\right)~.
\end{split}
\end{equation}

We evaluate the derivative by taking the symmetric difference around the point it is evaluated at. This has an error of $\mathcal{O}(1/N^2)$, so the error on that part of the Hamiltonian does not change orders of magnitude. We define $\Phi_{m,n}=\sqrt{\frac{\pi}{N}}\Phi_m(\theta_n)$ and $\Pi_{m,n}=\sqrt{\frac{\pi}{N}}\Pi_{m}(\theta_n)$. The commutation relations are now $[\Phi_{mn},\Pi_{m'n'}]=i\delta_{mm'}\delta_{nn'}$ and the above terms take the form:
\begin{eqnarray}
&&\frac{1}{2} \sum_{m=-\infty}^{\infty}\left(\frac{1}{2}\Pi_{m,1}^2+\sum_{n=2}^{N-2} \Pi_{mn} ^2 +\frac{1}{2}\Pi_{m,N-1}^2 \right) ~, \ \\
&&\frac{1}{2} \sum_{m=-\infty}^{\infty} \left(\sum_{n=1}^{N-2} \frac{N^2}{\pi^2} \sin \theta_{n+\frac{1}{2}} \left[\frac{\Phi_{m,n+1}}{\sin\theta_{n+1}}-\frac{\Phi_{m,n}}{\sin\theta_{n}} \right]^2\right) ~,
\end{eqnarray}
\begin{equation}
\begin{split}
\frac{1}{2} \sum_{m=-\infty}^{\infty}\left(\frac{1}{2}\left(\frac{m^2}{\sin^2\theta_1}+\mu^2\right)\Phi_{m,1}^2  + \sum_{n=2}^{N-2}  \left(\frac{m^2}{\sin^2\theta_n}+\mu^2\right)  \Phi_{m,n}^2 + \right. \\ \left. \frac{1}{2}\left(\frac{m^2}{\sin^2\theta_{N-1}}+\mu^2\right)\Phi_{m,N-1}^2\right)~.
\end{split}
\end{equation}
We make a final set of re-definitions to ensure we have both canonical commutation relations and properly scaled momenta in the Hamiltonian: $\tilde{\Pi}_{m,n}=\frac{1}{\sqrt{2}} \Pi_{m,n}$ and $\tilde{\Phi}_{m,n}=\sqrt{2} \Phi_{m,n}$ for $n=1$ and $n=N-1$. We omit the tilde for simplicity. The Hamiltonian is now:\footnote{Using the fact that $\frac{\sin\left(\frac{(n-1/2)\pi}{N}\right)+\sin\left(\frac{(n+1/2)\pi}{N}\right)}{\sin\left(\frac{n\pi}{N}\right)}=2\cos \frac{\pi}{2N}$.}
\begin{equation}
\begin{split}
H=\frac{1}{2}\sum_{m=-\infty}^{\infty} \left( \sum_{n=1}^{N-1} \Pi_{mn}^2+ \frac{N^2}{\pi^2} \left[ \frac{\sin\theta_{3/2}}{2\sin\theta_1} \Phi_{m,1}^2 + \frac{\sin\theta_{N-3/2}}{2\sin\theta_{N-1}} \Phi_{m,N-1}^2 + \sum_{n=2}^{N-2} 2\cos\left(\frac{\pi}{2N}\right)\Phi_{mn}^2 \right. \right.   \\ \left.
\left. -\frac{\sqrt{2}\sin\theta_{3/2}}{\sqrt{\sin\theta_1 \sin\theta_2}}\Phi_{m,1}\Phi_{m,2} - \frac{\sqrt{2}\sin\theta_{N-3/2}}{\sqrt{\sin\theta_{N-1} \sin\theta_{N-2}}}\Phi_{m,N-1}\Phi_{m,N-2} - 2 \sum_{n=2}^{N-3} \frac{\sin\theta_{n+1/2}}{\sqrt{\sin\theta_n \sin\theta_{n+1}}}\Phi_{m,n}\Phi_{m,n+1} \right] \right. \\
\left. + \frac{1}{4}\left(\frac{m^2}{\sin^2\theta_1}+\mu^2\right)\Phi_{m,1}^2  + \sum_{n=2}^{N-2}  \left(\frac{m^2}{\sin^2\theta_n}+\mu^2\right)  \Phi_{m,n}^2 + \frac{1}{4}\left(\frac{m^2}{\sin^2\theta_{N-1}}+\mu^2\right)\Phi_{m,N-1}^2 \right) ~.
\end{split}
\end{equation}

$H$ decouples into $H_m$'s that do not depend on the sign of $m$, so we can write:

\begin{equation}
H=H_0+2\sum_{m=1}^{\infty}H_m ~.
\label{hm}
\end{equation}

We can evaluate the contributions to entanglement entropy and mutual information coming from each of these $H_m$'s using the method outlined previously and obtain the total entanglement entropy with 
\begin{equation}
S=S_0+2\sum_{n=1}^\infty S_m ~.
\label{sm}
\end{equation}
When $|m|>N$, the diagonal terms in $H_m$ are generically larger than the off-diagonal ones. In other words, the $\Phi_{m,n}$ decouple at large $m$, which suggests that the sum (\ref{sm}) converges and therefore we can approximate it numerically by cutting it off at some $m_{max}=N^p$ for some power $p>1$. The appendix offers numerical evidence for this statement.  

\subsection{Fuzzy Sphere}
The noncommutative sphere is obtained by replacing Cartesian coordinates $x_i, i=1,2,3$ with 
\begin{equation}
X_i=R\frac{L_i}{\sqrt{J(J+1)}} ~,
\end{equation}
where $L_i$ are the generators of the $N=2J+1$-dimensional irreducible representation of $SU(2)$, i.e. $[L_i,L_j]=i\epsilon_{ijk} L_k$. 

\begin{figure}[t]
\centering
\includegraphics[scale=0.30]{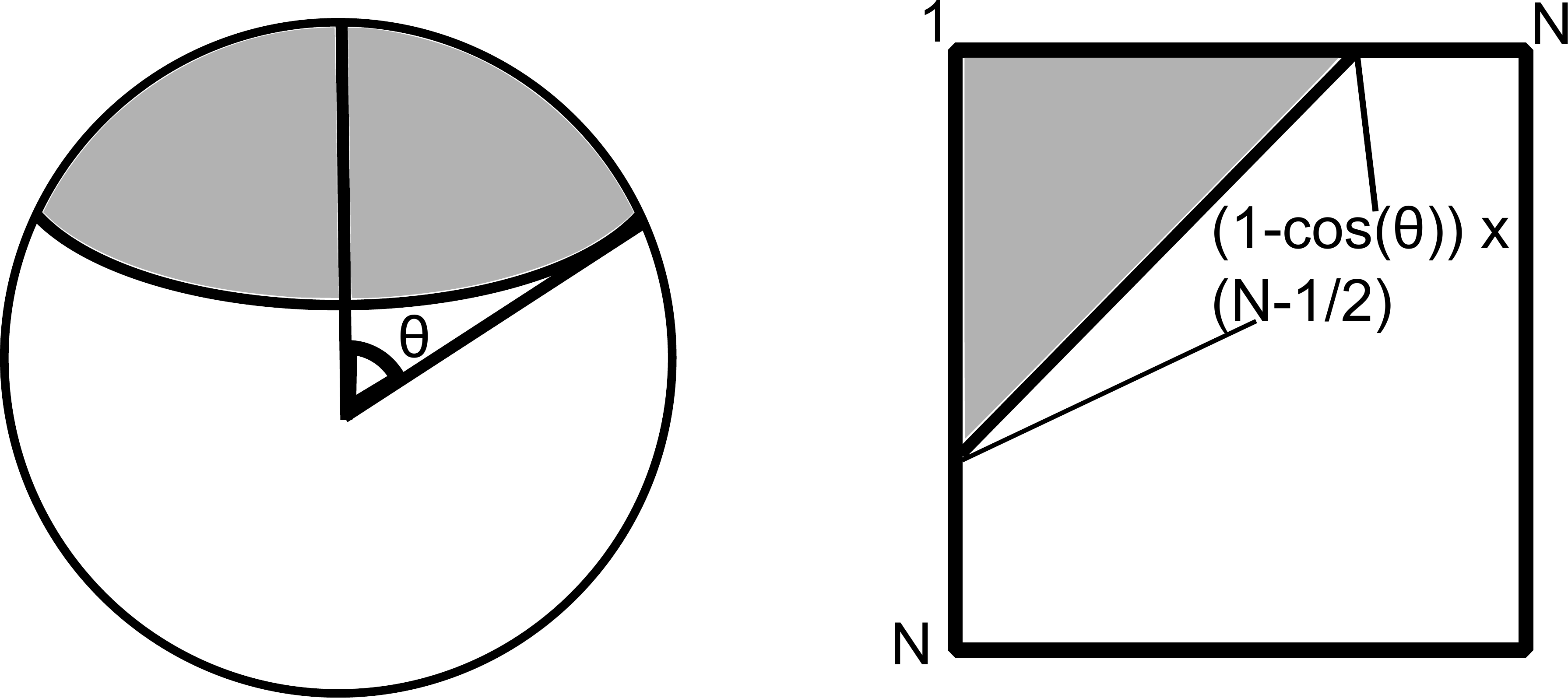}
\caption{Degrees of freedom on a fuzzy sphere and matching matrix components \cite{Karczmarek:2014}}
\label{dof}
\end{figure}

This can be motivated by the fact that $L_i L_i=J(J+1)\mathbb{I}$, so that $X_i X_i=R^2\mathbb{I}$, just like we have $x_ix_i=R^2$ for a commutative sphere \cite{Madore:1992, Douglas:2001ba}. A real scalar field on the fuzzy sphere corresponds to an $N \times N$ Hermitian matrix $\Phi$, and the Laplacian acting on the field is
\begin{equation}
-\frac{1}{R^2}[L_i,[L_i,\Phi]] ~,
\end{equation}
since the $L_i$ generate rotations. Integration on the fuzzy sphere is a trace
\begin{equation}
\frac{4 \pi R^2}{N} \text{Tr}(\cdot) ~,
\end{equation}
with the prefactor chosen so that the identity function maps to the unit matrix. 
The Hamiltonian for a free scalar field on the fuzzy sphere is then
\begin{equation}
H=\frac{4\pi R^2}{N} \frac{1}{2} \text{Tr}\left\{\dot{\Phi}^2-R^{-2}[L_i,\Phi]^2 + \mu^2 \Phi^2 \right\}~.
\end{equation} 
This is at most quadratic in every matrix element $[\Phi]_{ij}$, we can therefore in principle calculate the entanglement entropy between any subset of those and the rest using equation (\ref{eenum}). For example,\footnote{This form, introduced in \cite{Karczmarek:2014}, is intuitively clear but not very efficient numerically. \cite{Dou:2006ni} gives an equivalent but faster prescription.} by labeling the entries of $\Phi$ as
\begin{equation}
\Phi=\left( \begin{array}{ccccc}
\Phi_1 & \frac{\Phi_2+i\Phi_3}{\sqrt{2}} & \frac{\Phi_4+i\Phi_5}{\sqrt{2}} & \frac{\Phi_7+i\Phi_8}{\sqrt{2}} & \dots  \\
\frac{\Phi_2-i\Phi_3}{\sqrt{2}} & \Phi_6 & \frac{\Phi_9+i\Phi_{10}}{\sqrt{2}} & \dots & \dots \\
\frac{\Phi_4-i\Phi_5}{\sqrt{2}} & \frac{\Phi_9-i\Phi_{10}}{\sqrt{2}}  & \dots & \dots & \dots \\
\frac{\Phi_7-i\Phi_8}{\sqrt{2}} & \dots  & \dots & \dots & \dots \\
\dots & \dots & \dots & \dots & \dots 
\end{array} \right) 
\end{equation}
we can write the Hamiltonian in the form of (\ref{Hform}) with 
\begin{equation}
[K]_{ij} \sim - \frac{1}{2} \frac{\partial^2 \text{Tr}([L_i,\Phi]^2)}{\partial \Phi_i \partial \Phi_j}+\mu^2 \delta_{ij} ~.
\end{equation}
The only difficulty lies in identifying which subset of $\Phi_i$ correspond to field values in a certain geometrical domain. Fortunately, an answer was provided in \cite{Karczmarek:2014}: the degrees of freedom above the $k^{th}$ anti-diagonal correspond to a range of polar angles $[0,\theta]$ with
\begin{equation}
\cos \theta= 1 - \frac{k}{N-\frac{1}{2}}~,
\label{dofdistr}
\end{equation}
as illustrated in figure \ref{dof}. This boundary is of course not perfectly sharp: it has a thickness of $\mathcal{O}(\frac{R}{\sqrt{N}})$. It should be pointed out that the natural UV cutoff for the theory we are considering here is of $\mathcal{O}(\frac{R}{N})$ since there are $N^2$ degrees of freedom spread on an area of $4\pi R^2$. From now on, we set $R=1$ for the fuzzy sphere. Some possible implications of the difference between the width of the boundary and the UV cutoff were discussed in \cite{Karczmarek:2014}, but they do not concern us much here. 

\begin{figure}[t]
\centering
\includegraphics[scale=0.4]{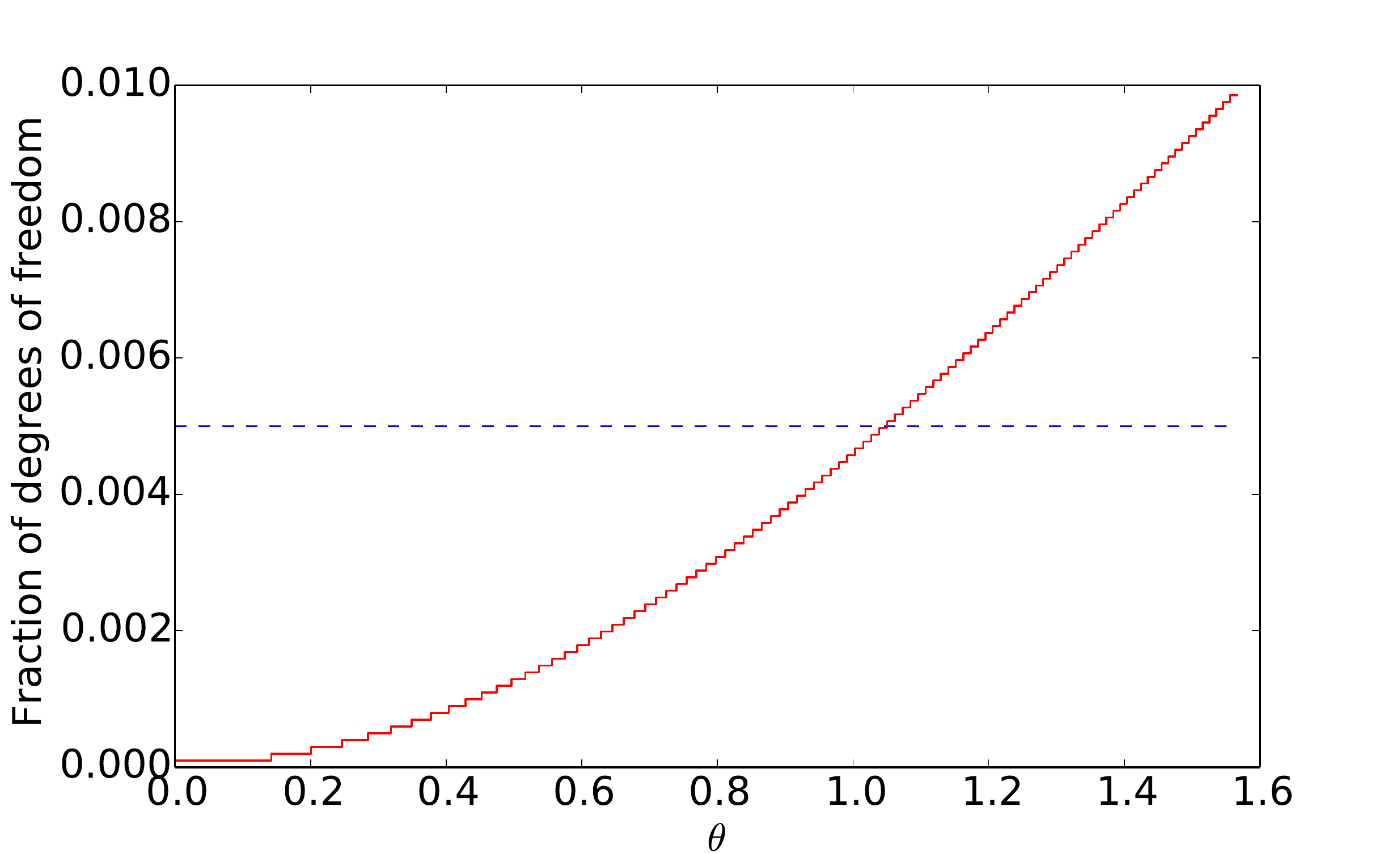}
\caption{Fraction of total number of degrees of freedom in a ring of constant area centered at polar angle $\theta$. The solid red line is for the fuzzy sphere at $N=100$, the dashed blue line corresponds to a fixed density.}
\label{dos}
\end{figure}

More potentially significant is the fact that the apparent distribution of UV degrees of freedom on the fuzzy sphere does not converge to that of the commutative sphere.  Using equation (\ref{dofdistr}), we can calculate that distribution on the fuzzy sphere along the polar direction by simply counting the number of elements in the corresponding part of the matrix. The fraction of matrix degrees of freedom corresponding to field degrees of freedom in a polar cap of area $A$ grows roughly as $A^2$, in contrast to the constant density of degrees of freedom on a commutative sphere. We can examine this in more detail by looking at rings centered at varying polar angles $\theta$. Consider two polar caps described by matrix degrees of freedom corresponding to triangles ending at $k$ and $k+1$. These caps have an area of $2\pi k/N$ and $2 \pi (k+1)/N$ respectively, making the area of the ring formed by removing the smaller cap from the larger cap $2\pi/N$ (independent of the position of the ring). But the value of the field on this ring is represented in the matrix by the $k$ degrees of freedom in the $k^{th}$ anti-diagonal line. Therefore, the fraction of degrees of matrix degrees of freedom describing a ring centered on $\theta$ is $\frac{k}{N(N+1)}=\frac{1-\cos\theta}{N+1}$. However, on a regularized commutative theory we would expect a fixed density of degrees of freedom per unit area. These two distributions are illustrated in figure \ref{dos}. It is interesting that, despite this apparent difference between the two theories, the UV-finite mutual information is the same, as we will show.

\section{Results}

\subsection{Entanglement entropy}
\begin{figure}[t]
\centering
\begin{subfigure}[b]{.4\textwidth} 
\centering
\includegraphics[scale=0.4]{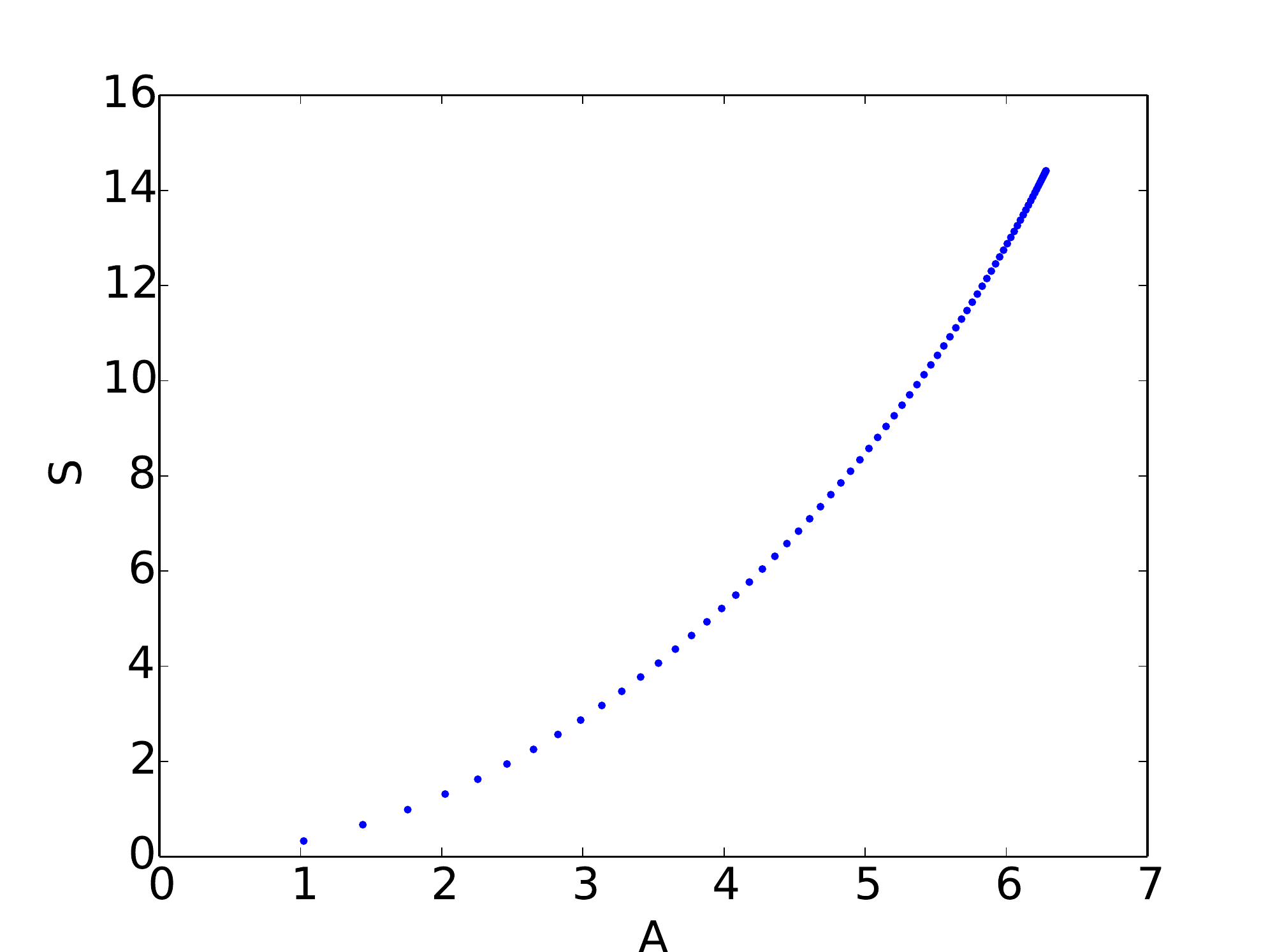}
\caption{Fuzzy sphere}
\label{sanc75}
\end{subfigure}\hfill
\begin{subfigure}[b]{.4\textwidth}
\centering
\includegraphics[scale=0.4]{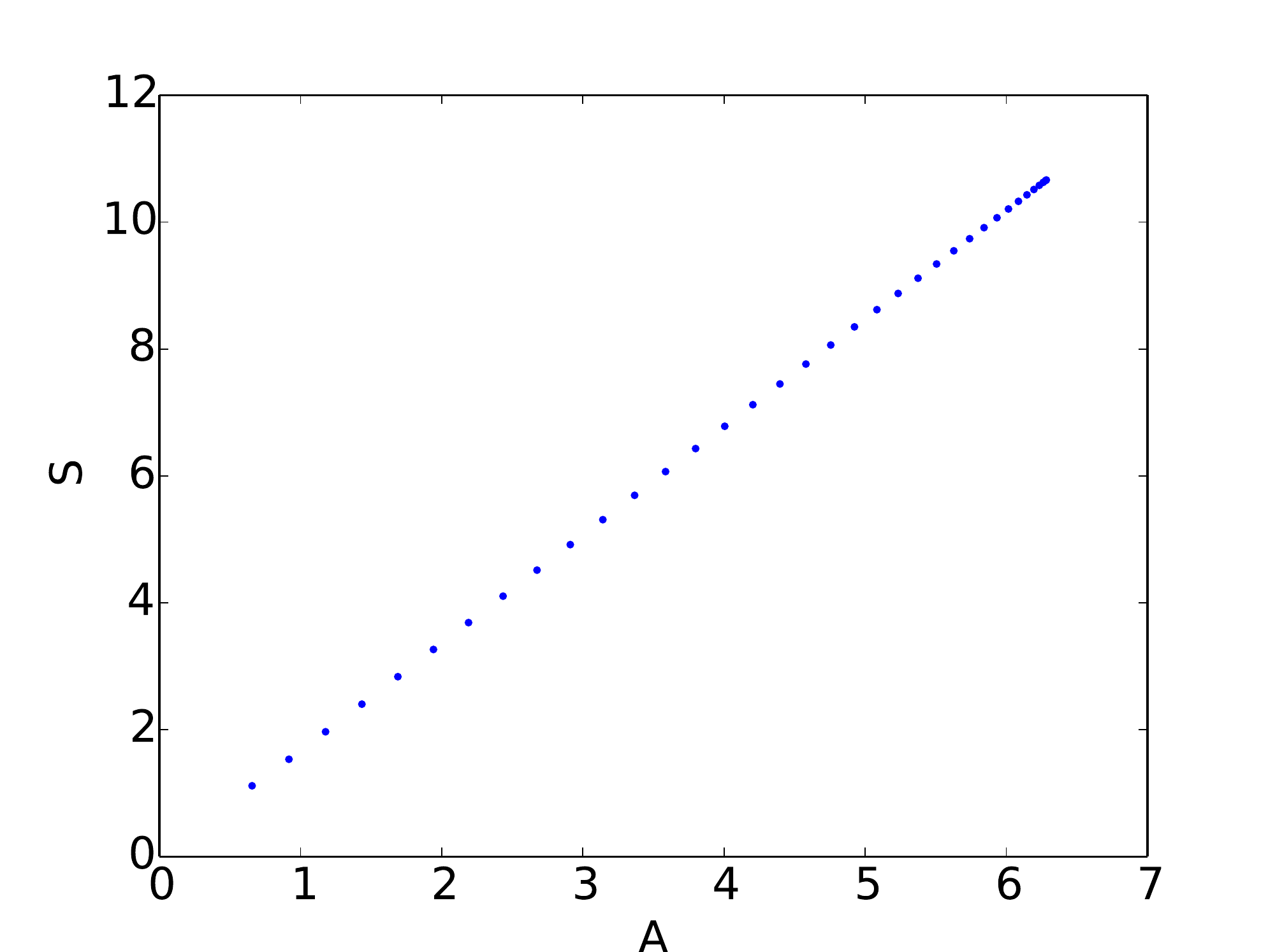}
\caption{Commutative sphere}
\label{sa75}
\end{subfigure}  
\caption{Entanglement entropy vs. area of boundary on a commutative sphere with $N=75$ for $\mu=1.0$.}
\end{figure}

We start by calculating the entanglement entropy of a polar cap for caps of varying size. In \cite{Karczmarek:2014}, it was shown that the entanglement entropy on a fuzzy sphere was not proportional to the length of the boundary (i.e. an ``area'' law) but was extensive for small regions and sub-extensive for larger ones, as seen in figure \ref{sanc75}. In contrast, the same quantity on the commutative sphere behaves as we would expect it to, as seen in figure \ref{sa75}. We can clearly see that the relation between the length of the boundary and the entropy is linear, with a very small y-intercept that is an artifact of discretization:
\begin{equation}
S_\text{comm}\approx aA ~.
\end{equation}

\begin{figure}[t]
\centering
\includegraphics[scale=0.4]{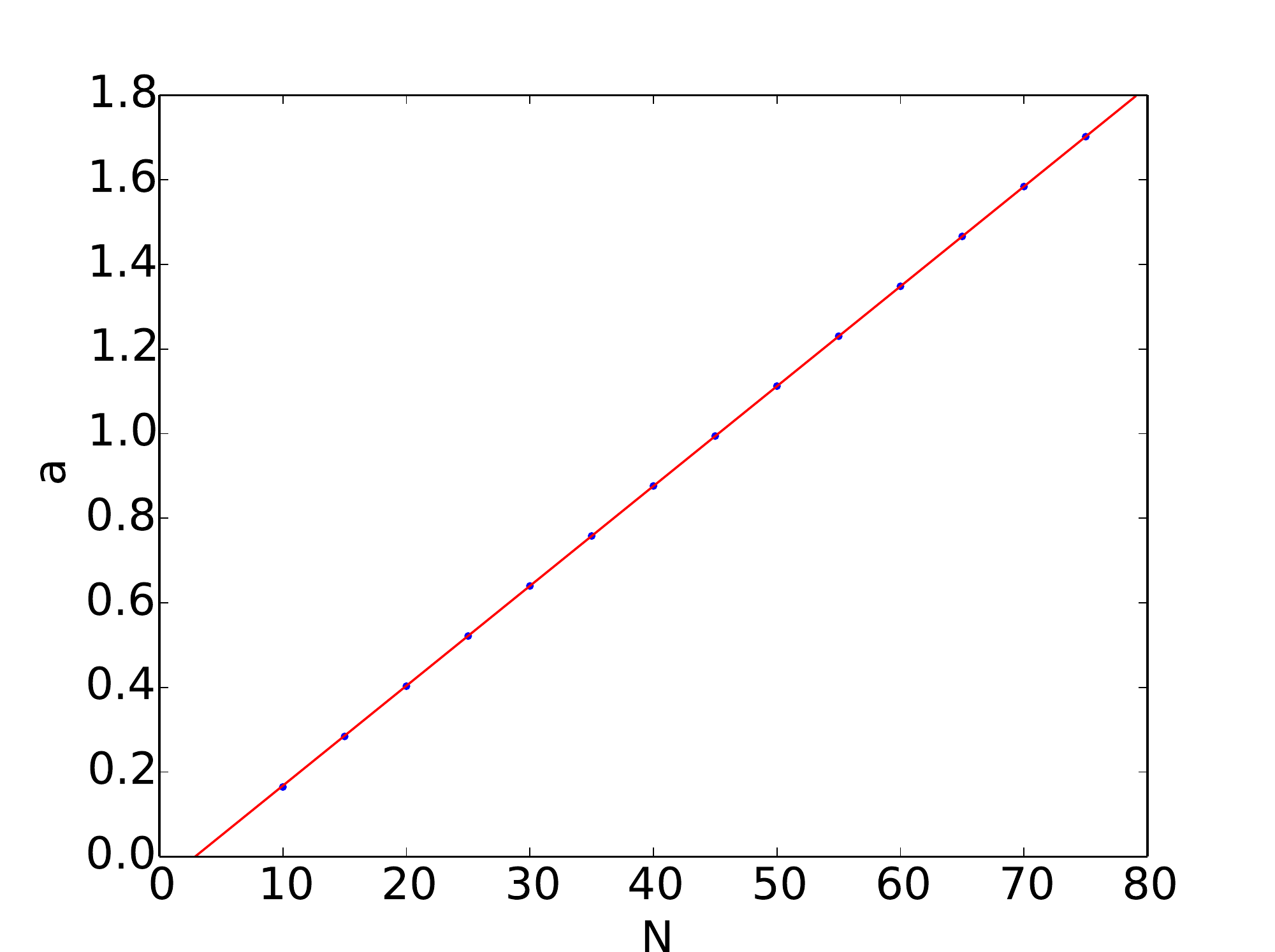}
\caption{Slope $a$ of $S_\text{comm}$ vs. $A$ at different N, with $\mu=1.0$ on the commutative sphere. The red fit line takes the form: $a=0.0236N-0.0682$}
\label{abvsn:a}
\end{figure}

We can study the parameter $a$ as a function of $N$, as seen in figure \ref{abvsn:a}. We can see that it has a term linear in N, as expected, and a constant term: $a=a_1 N + a_2$. Therefore, we can write
\begin{equation}
S_\text{comm}=\alpha \frac{A}{\epsilon} + \beta A  + \cdots ~,
\label{S}
\end{equation} 
where `$\cdots$' stands for terms that go to zero as $\epsilon \rightarrow 0$. Since $\epsilon=\frac{\pi}{N}$, our fit tells us that $\alpha=0.074$ and $\beta=-0.068$. 

\subsection{Mutual information}
\begin{figure}[t]
\begin{subfigure}[b]{0.4\textwidth}
\includegraphics[scale=0.3]{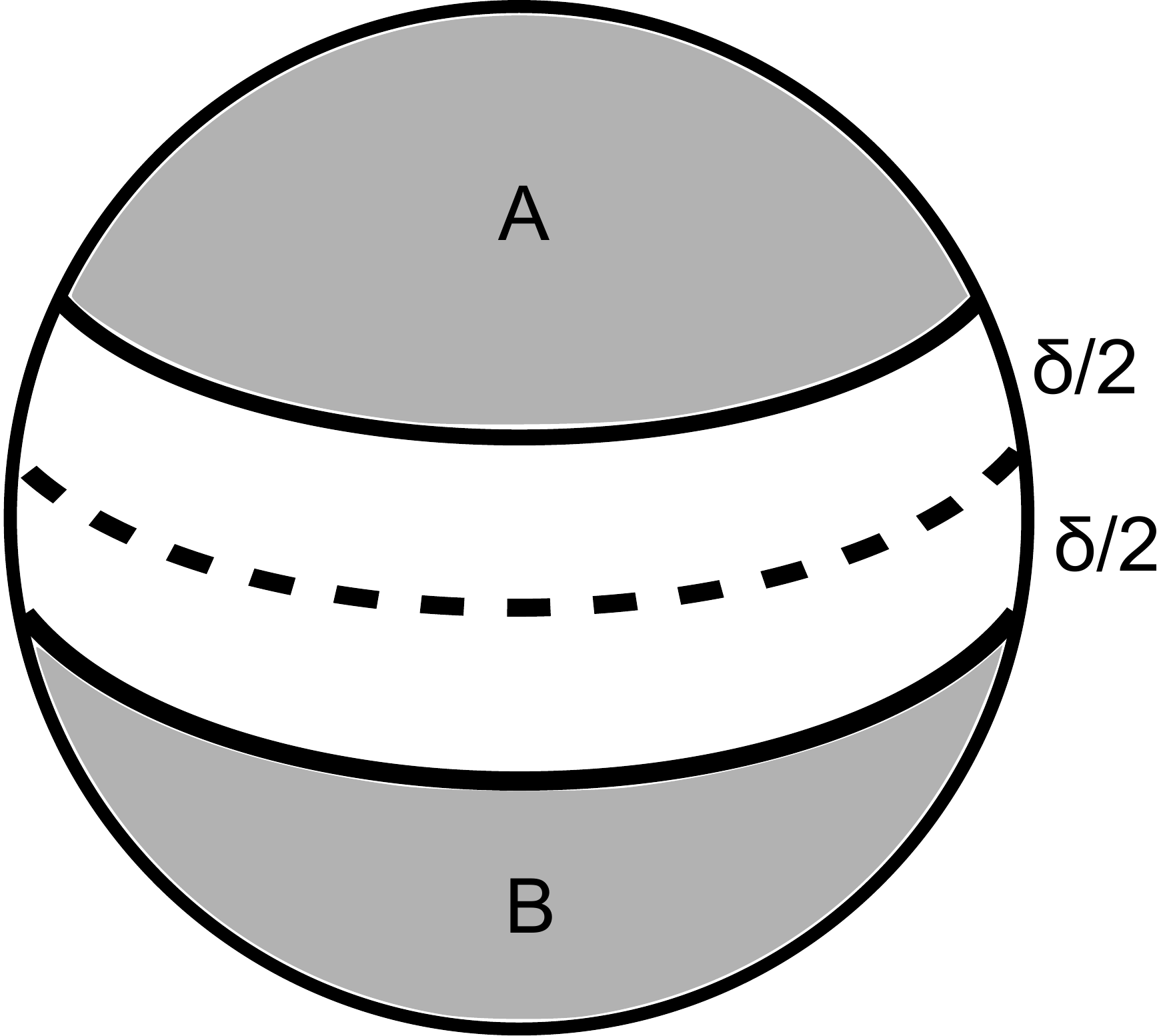}
\caption{Regions on the sphere}
\label{setsphere}
\end{subfigure} \hfill
\begin{subfigure}[b]{0.4\textwidth}
\includegraphics[scale=0.3]{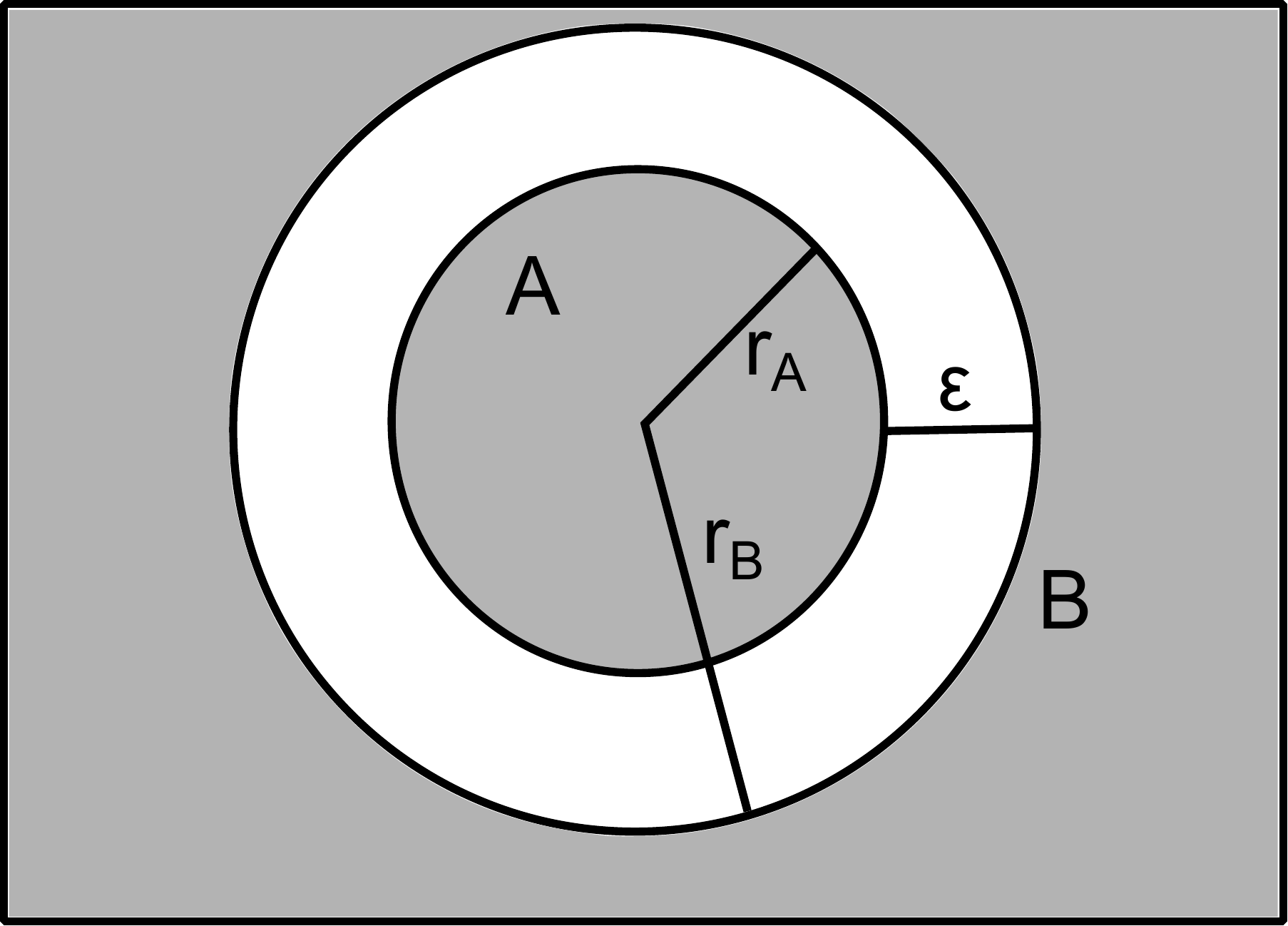}
\caption{Regions conformally mapped to the plane}
\label{setplane}
\end{subfigure}
\caption{Regions A and B between which we calculate the mutual information $I(A,B)$.}
\end{figure}

\begin{figure}[t]
\centering
\includegraphics[scale=0.4]{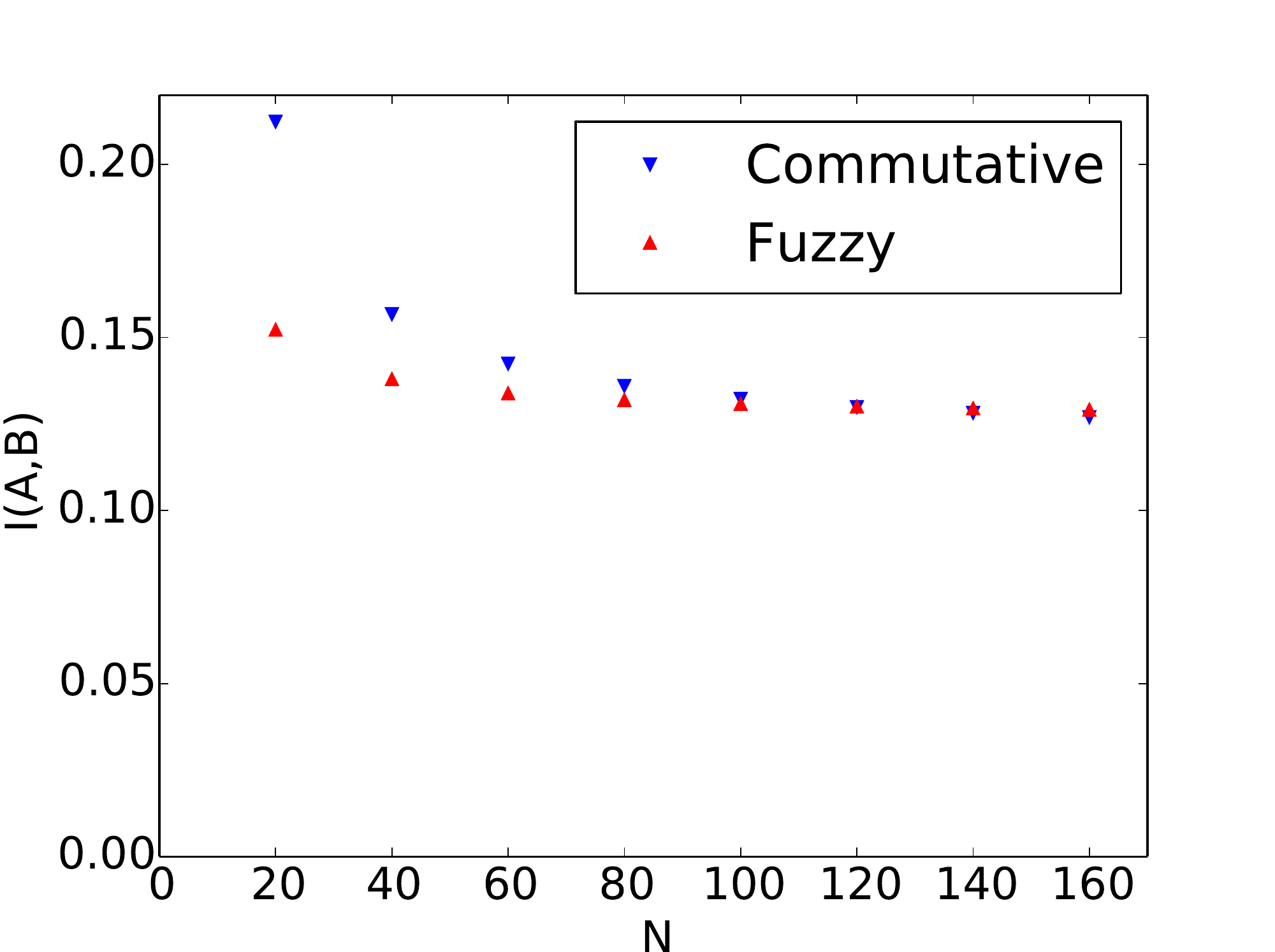}
\caption{Mutual information for two polar caps separated by an annulus centered on the equator. On the commutative sphere, the annulus has a width of 0.628 rad and I(A,B) goes to 0.12 faster than $\frac{1}{N}$. On the fuzzy sphere, the annulus has a width off 0.609 rad and I(A,B) it goes to 0.13 faster than $\frac{1}{N}$.}
\label{halfi}
\end{figure}

\begin{figure}[t]
\centering
\includegraphics[scale=0.35]{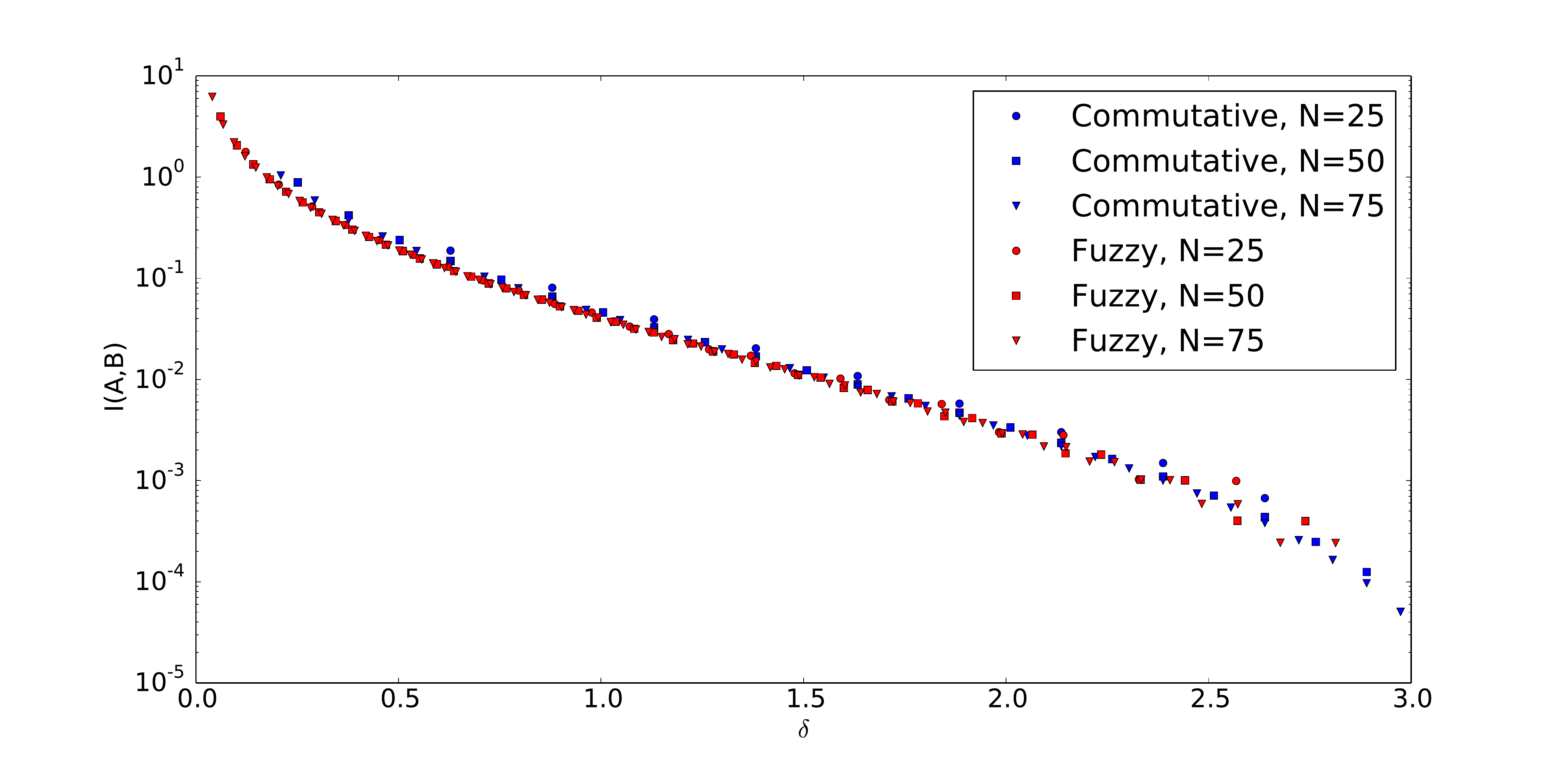}
\caption{Mutual information for two polar caps separated by an annulus of angular width $\delta$ centered on the equator, $\mu=1.0$}
\label{halfcnc}
\end{figure}

The easiest UV-finite quantity to calculate from entanglement entropy is the mutual information between two polar caps separated by an annulus centered on the equator with width $\delta$ (see figure \ref{setsphere}). Figure \ref{halfi} shows the result of this calculation for a fixed angular separation of about $0.2\pi$:\footnote{Because of the differences in regularization, polar caps on the commutative and fuzzy spheres do not actually have their boundaries at the same $\theta$. We have picked here polar caps that are separated by $0.2\pi \approx 0.628$ for the commutative sphere and the closest possible value on the non-commutative sphere: $\theta \approx 0.609$} it is easy to see that it asymptotes to a finite value as $N$ is increased for both the commutative and fuzzy spheres and that the value for both cases is similar.  They appear to be consistent. We can repeat the calculation for various widths of the central annulus. This is shown in figure \ref{halfcnc}, where we can also see the convergence to a finite value as $N$ increases.

The $\delta \sim 0$ and $\delta \sim \pi$ regions can be studied analytically for a conformally coupled theory, i.e. for $\mu^2=\frac{1}{8} \mathcal{R}$, where $\mathcal{R}=\frac{2}{R^2}=2.0$ is the curvature of the sphere. Following the argument in appendix A of \cite{Herzog:2014} and using results from \cite{Casini:2011} , we note that $\mathbb{R}\times S^2$ is conformally related to $\mathbb{R}^{1,2}$ by

\begin{equation}
t\pm r = \tan \left( \frac{\tau \pm \theta}{2} \right), \phi=\phi ~,
\end{equation}
where the left-hand side coordinates are those on the plane and the right-hand side ones are those on the sphere. An entangling surface $\tau=0, ~ \theta=\theta_0$ corresponds to a circle centered at zero with radius $\tan{\theta_0/2}$. 
 Polar caps with $\theta=\frac{\pi \pm \delta}{2}$ are then mapped to concentric disks on the plane centered at zero with radii $r_{1,2}=\tan \left(\frac{\pi \pm \delta}{4} \right)$. The only conformally invariant quantity that can be constructed from geometrical data on these two disks is the cross-ratio
\begin{equation}
x=\frac{4 r_1 r_2}{(r_1-r_2)^2}=\cot^2 (\delta /2)  ~.
\end{equation}
Since mutual information is invariant under conformal transformations, it must have an expansion in powers of $x$. It was shown in \cite{Cardy:2013} that as $x \rightarrow 0$ (i.e. in the region where $\delta \sim \pi$) the mutual information takes the form of
\begin{equation}
I(\delta)= \frac{1}{12} x + \mathcal{O}(x^2) \approx \frac{1}{12} \cot^2 \frac{\delta}{2} ~.
\label{larged}
\end{equation}    
The $\delta \sim 0$ behaviour can be obtained by looking at the limit when $|r_1-r_2|\rightarrow 0$ and matching to the area law \cite{Herzog:2014}. Using the result in \cite{Casini:2009}, we know that
\begin{equation}
I(\delta) \approx 0.0397 \frac{A}{\varepsilon} ~,
\end{equation}
where $A$ is the length of the boundary between the two disks in flat space and $\varepsilon$ is the distance between them (see figure \ref{setplane}). We take $A=2\pi \sqrt{r_1 r_2}$ (the geometric mean of the boundary lengths) and $\varepsilon=|r_1-r_2|$ to obtain \cite{Herzog:2014}
\begin{equation}
I(\delta) \approx 0.0397 \cdot 2\pi \frac{\sqrt{r_1 r_2}}{|r_1-r_2|} \approx 0.125 \cot \frac{\delta}{2} ~.
\label{smalld}
\end{equation}

\begin{figure}[t]
\centering
\includegraphics[scale=0.35]{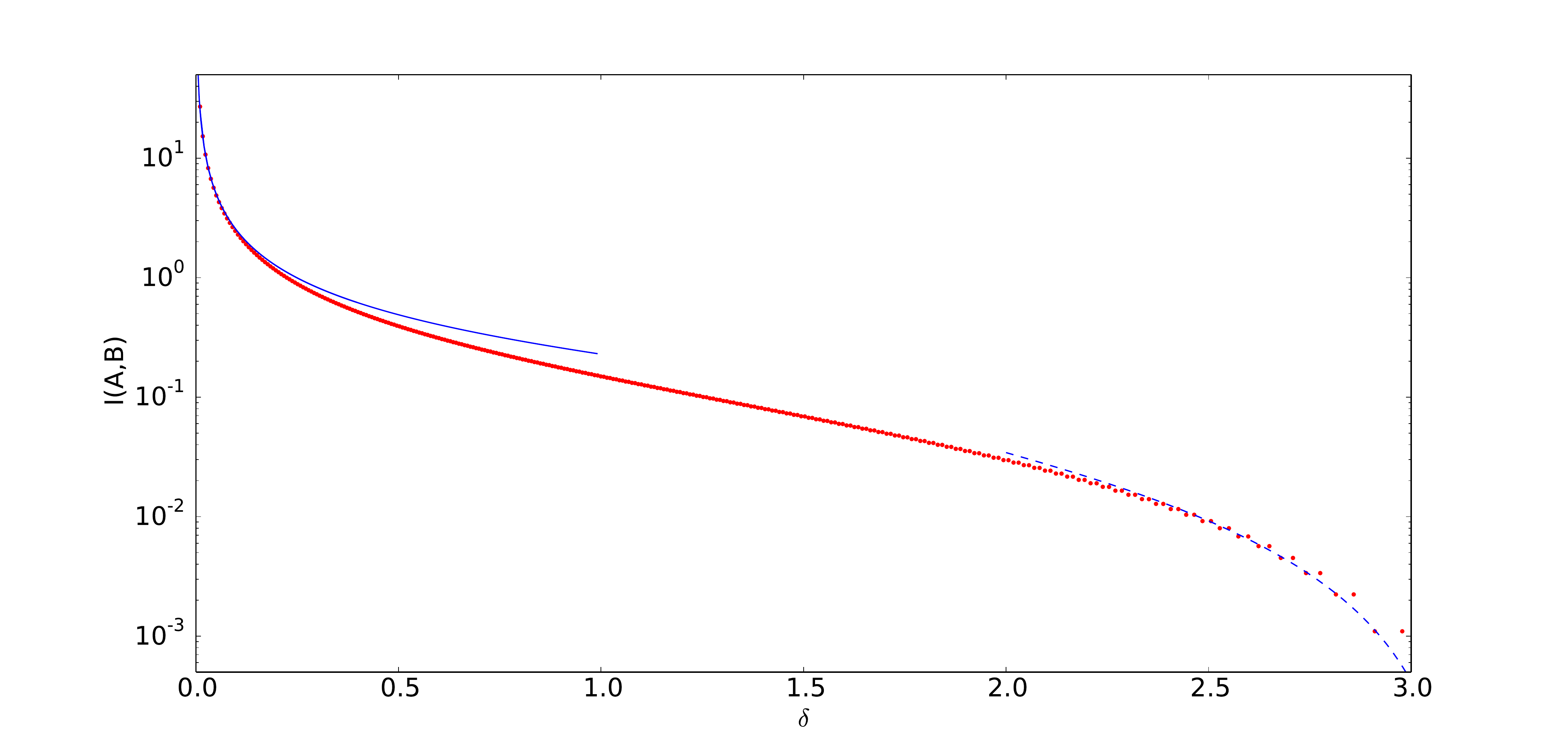}
\caption{Mutual information on the fuzzy sphere for two polar caps separated by an annulus of angular width $\delta$ centered on the equator. Calculation done at conformal coupling ($\mu=0.5$) with N=300. The solid and dashed lines correspond to the analytical predictions (\ref{smalld}) and (\ref{larged}) for a commutative sphere at small and large $\delta$ respectively.}
\label{infoconf}
\end{figure}

Figure \ref{infoconf} shows the mutual information for a conformal scalar on the fuzzy sphere at a high value of N, as well as curves for both the small and large $\delta$ behaviour of a conformal scalar on the commutative sphere. We can see that these agree when we expect them to. 

\begin{figure}[t]
\centering
\includegraphics[scale=0.35]{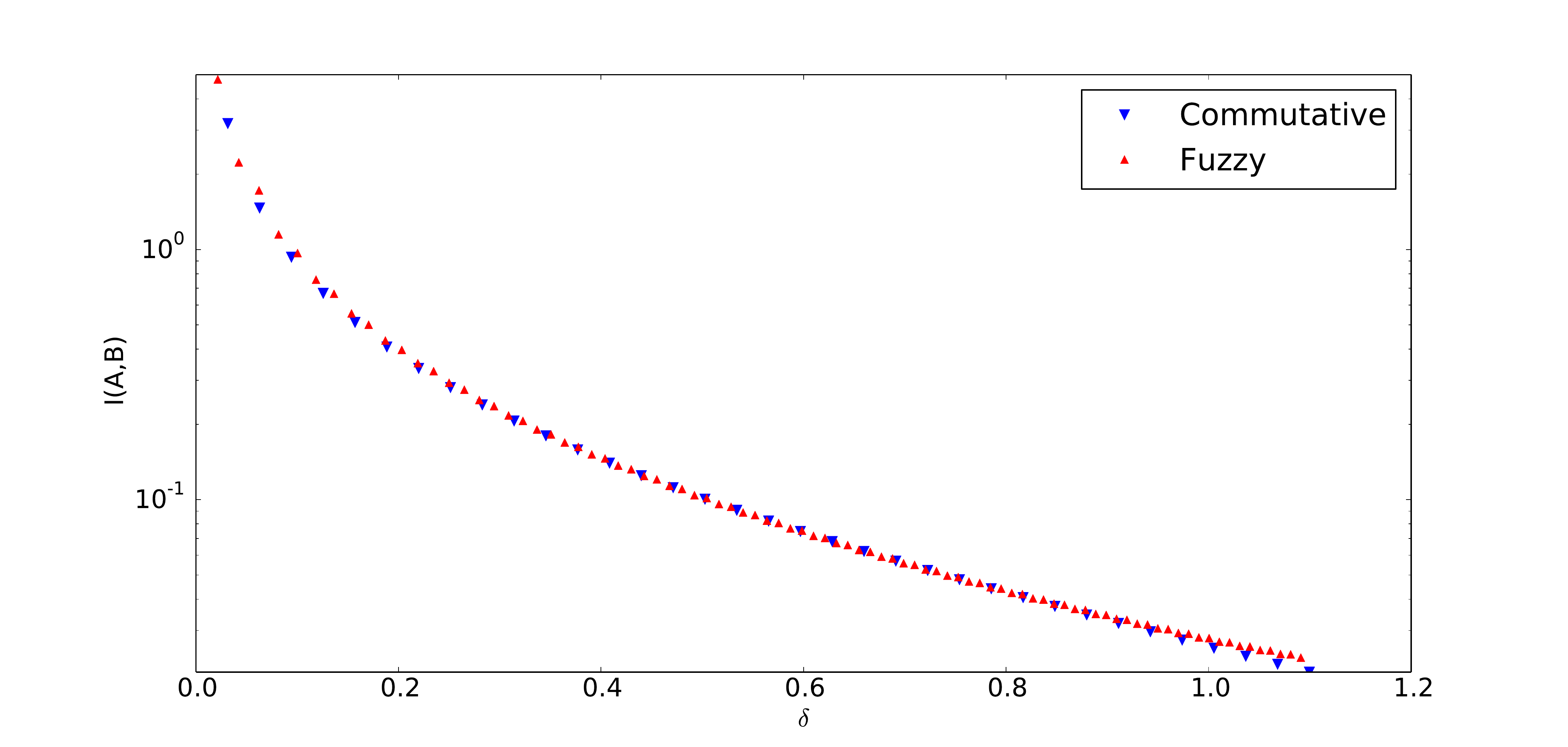}
\caption{Mutual information between a polar cap $A$ of fixed size and a region $B$ separated by $A$ by an annulus of size $\delta$. On the commutative sphere , $A$ terminates at $\theta\approx0.44 \; \text{rad}$ and on the fuzzy sphere it terminates at $\theta\approx0.45 \; \text{rad}$. In both cases,$N=100$ and $\mu=1.0$.}
\label{offcentre}
\end{figure}

To ensure that the symmetry of the previous setup does not lead to unusual cancellations, we can consider more generic regions $A$ and $B$. The most convenient configuration is to fix the size of $A$ and vary the width $\delta$ of the annulus. The results for a particular size of $A$ are shown in figure \ref{offcentre}: we can again see a striking agreement between the commutative and non-commutative theories. 

\pagebreak
\section{Discussion and Outlook}
In this paper, we have calculated the entanglement entropy for free scalar fields on a sphere. We have confirmed the expected area-law UV divergence even when the theory is not conformal, and reiterated that this area law is not followed by the noncommutative theory. We have also seen that the mutual information is the same for the commutative and noncommutative spheres. We calculated this not only in the case of conformal coupling, where we could make an analytical prediction, but also at non-conformal mass. There are two things to note here. First, the mutual information for a non-conformal scalar field behaves qualitatively as the mutual information for the very specific case of a conformal field. More importantly, mutual information on the fuzzy sphere matches that on the commutative sphere. This validates the approach, first taken in \cite{Karczmarek:2014}, of identifying upper-left triangles of the field matrix with the value of the field on polar caps (as in figure \ref{dof}). While this approach appears to create an exotic distribution of degrees of freedom (as described at the end of section 2.2), the behaviour of low-energy modes is unaffected, at least for free field theories. It would be interesting to understand this further. In particular, it would be interesting to understand how the IR degrees of freedom arise from the matrix model. 

In light of our result, the differing behaviour of mutual information seen in \cite{Karczmarek:2013} cannot be solely attributed to noncommutativity. Instead, it is likely caused by a combination of noncommutativity, strong coupling and large $N_c$. A natural extension of this present work is therefore to find a way to repeat the calculation of mutual information in a coupled non-commutative field theory to see if that is enough to create a change.

One could also take as a starting point the theory with conformal mass and treat the non-conformal mass as a perturbation to a CFT in order to calculate analytically mutual information for the non-conformal theory, using the framework presented in \cite{Rosenhaus:2014woa}.\footnote{I thank Aitor Lewkowycz for bringing this to my attention.} This approach has been taken to analytically  study renormalized entanglement entropy (REE) for a scalar field on a commutative plane \cite{Lee:2014zaa}. The influence of relevant perturbations on REE for a scalar field on a commutative sphere has also been studied numerically \cite{Klebanov:2012va, Nishioka:2014kpa}. It would be interesting to compare REE on a noncommutative sphere to that on an ordinary sphere, given that we found large differences in the UV but agreement in the IR.

Finally, it could be instructive to look at mutual information in other non-local theories where entanglement entropy is known to violate the area law, such as in \cite{Shiba:2013jja} or \cite{Pang:2014tpa}.

\acknowledgments 
I am grateful to Joanna Karczmarek for originally proposing this project as well as for numerous discussions, continuing guidance and advice on the manuscript. I am also thankful to Charles Rabideau for general discussions and for suggesting the non-symmetric setup described at the end of section 3.2. This work is supported in part by an award from the Fonds de Recherche du Qu\'ebec --- Nature et Technologies (FRQNT).


\appendix

\section{Convergence}
\begin{figure}[t]
\begin{subfigure}[b]{0.4\textwidth}
\includegraphics[scale=0.3]{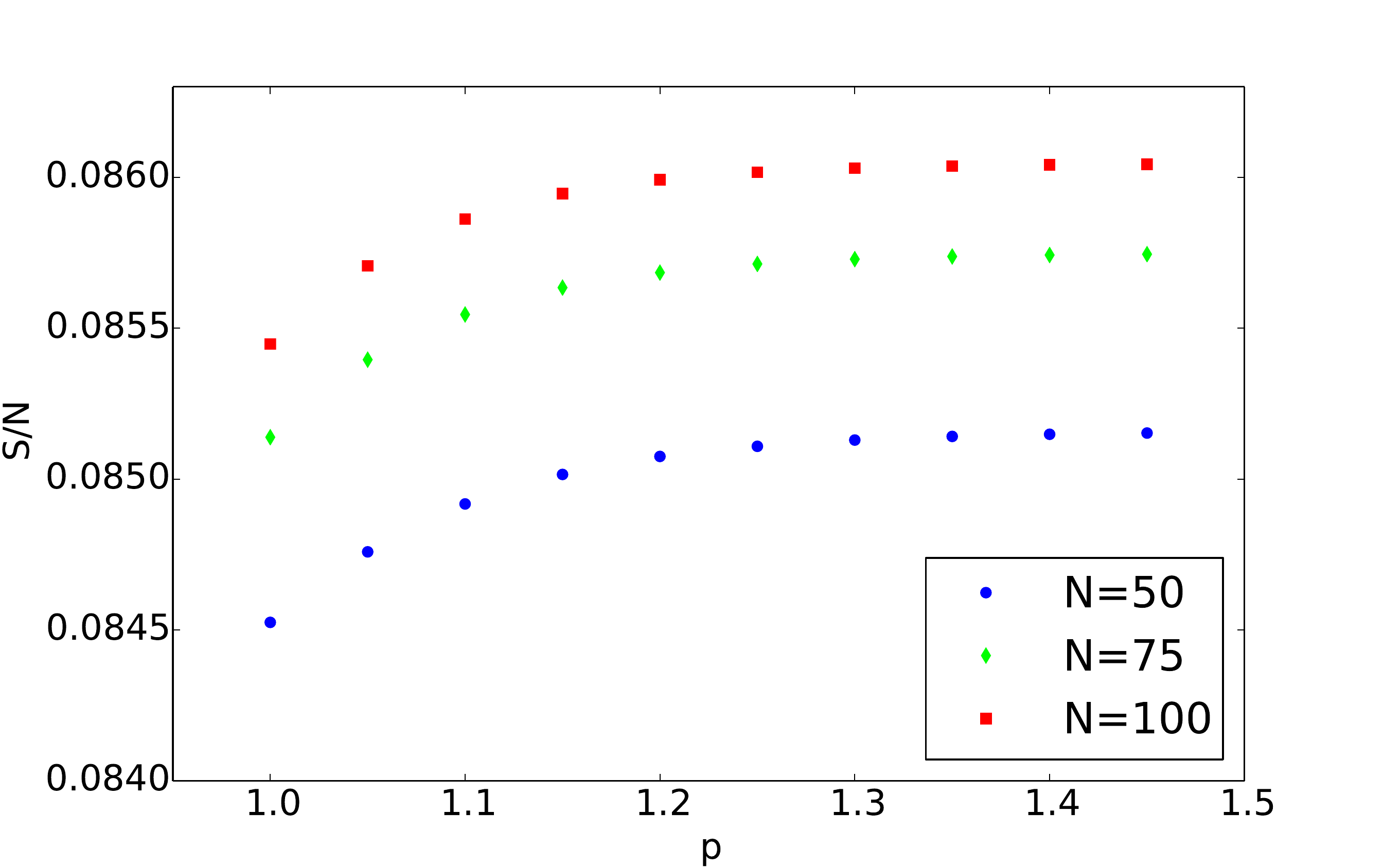}
\caption{Small polar cap: $\theta \approx \arccos(0.8)$}
\end{subfigure} \hfill
\begin{subfigure}[b]{0.4\textwidth}
\includegraphics[scale=0.3]{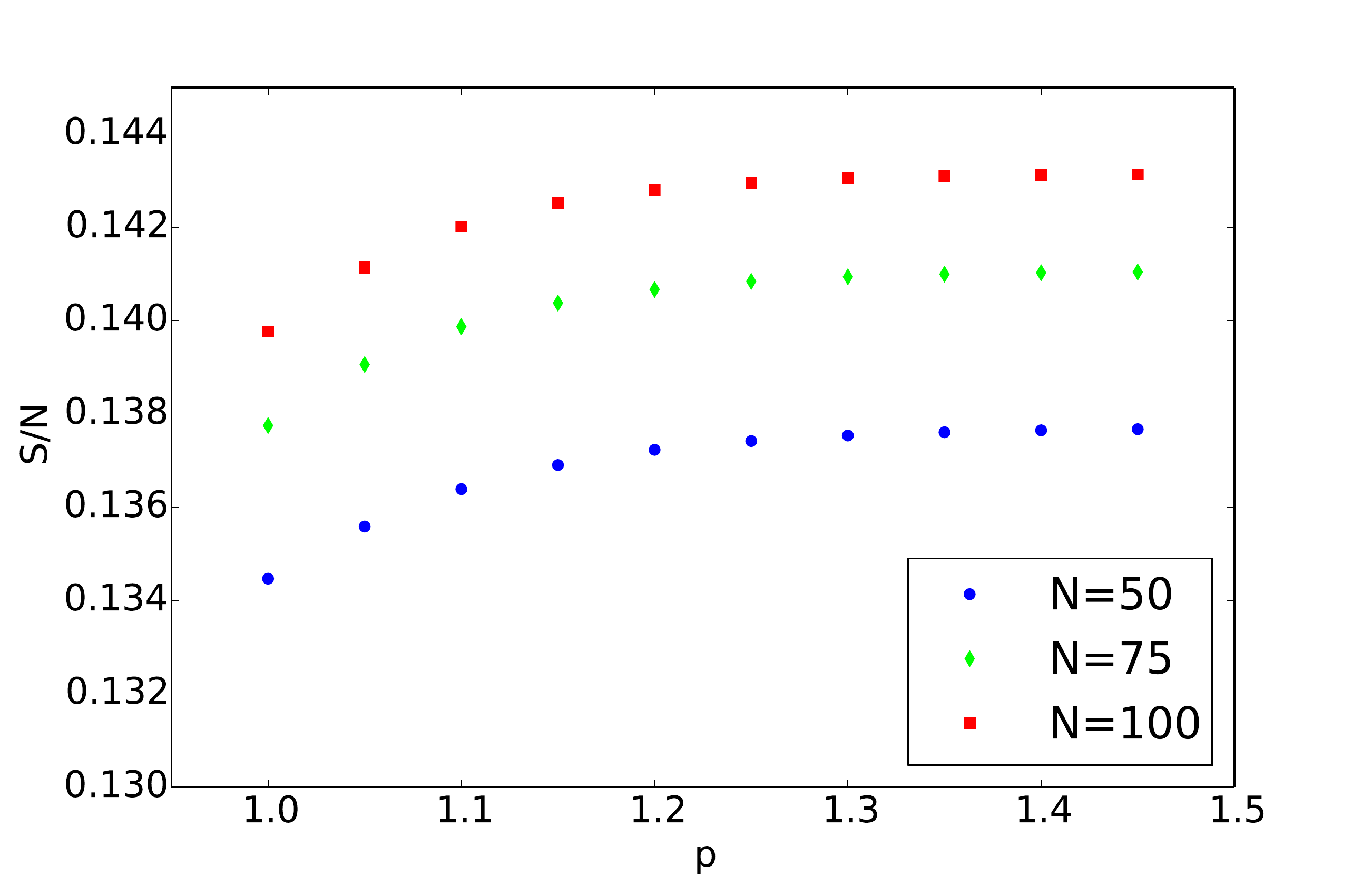}
\caption{Large polar cap: $\theta \approx \frac{\pi}{2}$}
\end{subfigure}
\caption{Scaled entanglement entropy for $\mu=1$ at different N as a function of the power $p$ of the cutoff $m_{max}=N^p$}
\label{stab}
\end{figure}

The infinite sum over azimuthal Fourier modes in (\ref{sm}) can be shown to converge numerically. To do so, we compute the entanglement entropy for a polar cap of small size ($\theta \approx 35^{\circ}$ and large size ($\theta \approx 90^{\circ}$) for different maximal m of the form $m_{max}=N^p$ (at different N). The results are shown in figure \ref{stab}. At $m\sim N^{4/3}$, the result differs from the asymptotic value by less than 0.05\%.


\bibliographystyle{JHEP}
\bibliography{MutualInfo}

\providecommand{\href}[2]{#2}\begingroup\raggedright\begin{thebibliography}{10}

\bibitem{Ryu:2006bv}
S.~Ryu and T.~Takayanagi, {\it {Holographic derivation of entanglement entropy
  from AdS/CFT}},  {\em Phys.Rev.Lett.} {\bf 96} (2006) 181602,
  [\href{http://xxx.lanl.gov/abs/hep-th/0603001}{{\tt hep-th/0603001}}].

\bibitem{Faulkner:2013}
T.~Faulkner, M.~Guica, T.~Hartman, R.~C. Myers, and M.~Van~Raamsdonk, {\it
  {Gravitation from Entanglement in Holographic CFTs}},
  \href{http://xxx.lanl.gov/abs/1312.7856}{{\tt arXiv:1312.7856}}.

\bibitem{Swingle:2014}
B.~Swingle and M.~Van~Raamsdonk, {\it {Universality of Gravity from
  Entanglement}},  \href{http://xxx.lanl.gov/abs/1405.2933}{{\tt
  arXiv:1405.2933}}.

\bibitem{Srednicki:1993im}
M.~Srednicki, {\it {Entropy and area}},  {\em Phys.Rev.Lett.} {\bf 71} (1993)
  666--669, [\href{http://xxx.lanl.gov/abs/hep-th/9303048}{{\tt
  hep-th/9303048}}].

\bibitem{Eisert:2008ur}
J.~Eisert, M.~Cramer, and M.~Plenio, {\it {Area laws for the entanglement
  entropy - a review}},  {\em Rev.Mod.Phys.} {\bf 82} (2010) 277--306,
  [\href{http://xxx.lanl.gov/abs/0808.3773}{{\tt arXiv:0808.3773}}].

\bibitem{Groisman:2005}
B.~Groisman, S.~Popescu, and A.~Winter, {\it {On the quantum, classical and
  total amount of correlations in a quantum state}},  {\em Phys. Rev. A} {\bf
  72} (2005) 032317, [\href{http://xxx.lanl.gov/abs/quant-ph/0410091}{{\tt
  quant-ph/0410091}}].

\bibitem{Fischler:2013gsa}
W.~Fischler, A.~Kundu, and S.~Kundu, {\it {Holographic Entanglement in a
  Noncommutative Gauge Theory}},  \href{http://xxx.lanl.gov/abs/1307.2932}{{\tt
  arXiv:1307.2932}}.

\bibitem{Karczmarek:2013}
J.~L. Karczmarek and C.~Rabideau, {\it {Holographic entanglement entropy in
  nonlocal theories}},  \href{http://xxx.lanl.gov/abs/1307.3517}{{\tt
  arXiv:1307.3517}}.

\bibitem{Karczmarek:2014}
J.~L. Karczmarek and P.~Sabella-Garnier, {\it {Entanglement entropy on the
  fuzzy sphere}},  {\em JHEP} {\bf 03} (2014) 129,
  [\href{http://xxx.lanl.gov/abs/1310.8345}{{\tt arXiv:1310.8345}}].

\bibitem{Dou:2006ni}
D.~Dou and B.~Ydri, {\it {Entanglement entropy on fuzzy spaces}},  {\em
  Phys.Rev.} {\bf D74} (2006) 044014,
  [\href{http://xxx.lanl.gov/abs/gr-qc/0605003}{{\tt gr-qc/0605003}}].

\bibitem{Dou:2009cw}
D.~Dou, {\it {Comments on the Entanglement Entropy on Fuzzy Spaces}},  {\em
  Mod.Phys.Lett.} {\bf A24} (2009) 2467--2480,
  [\href{http://xxx.lanl.gov/abs/0903.3731}{{\tt arXiv:0903.3731}}].

\bibitem{Nishioka:2009un}
T.~Nishioka, S.~Ryu, and T.~Takayanagi, {\it {Holographic Entanglement Entropy:
  An Overview}},  {\em J.Phys.} {\bf A42} (2009) 504008,
  [\href{http://xxx.lanl.gov/abs/0905.0932}{{\tt arXiv:0905.0932}}].

\bibitem{Casini:2009}
H.~Casini and M.~Huerta, {\it {Entanglement entropy in free quantum field
  theory}},  {\em J. Phys.} {\bf A42} (2009) 504007,
  [\href{http://xxx.lanl.gov/abs/0905.2562}{{\tt arXiv:0905.2562}}].

\bibitem{Herzog:2014}
C.~P. Herzog, {\it {Universal Thermal Corrections to Entanglement for Conformal
  Field Theories on Spheres}},  \href{http://xxx.lanl.gov/abs/1407.1358}{{\tt
  arXiv:1407.1358}}.

\bibitem{Madore:1992}
J.~Madore, {\it {The fuzzy sphere}},  {\em Class. Quant. Gravity} {\bf 9}
  (1992) 69--88.

\bibitem{Douglas:2001ba}
M.~R. Douglas and N.~A. Nekrasov, {\it {Noncommutative field theory}},  {\em
  Rev.Mod.Phys.} {\bf 73} (2001) 977--1029,
  [\href{http://xxx.lanl.gov/abs/hep-th/0106048}{{\tt hep-th/0106048}}].

\bibitem{Casini:2011}
H.~Cassini, M.~Huerta, and R.~C. Myers, {\it {Towards a derivation of
  holographic entanglement entropy}},  {\em JHEP} {\bf 1105} (2011) 036,
  [\href{http://xxx.lanl.gov/abs/1102.0440}{{\tt arXiv:1102.0440}}].

\bibitem{Cardy:2013}
J.~Cardy, {\it {Some Results on Mutual Information of Disjoint Regions in
  Higher Dimensions}},  {\em J. Phys.} {\bf A: Math. Theor. 46} (2013) 5285402,
  [\href{http://xxx.lanl.gov/abs/1304.7985}{{\tt arXiv:1304.7985}}].

\bibitem{Rosenhaus:2014woa}
V.~Rosenhaus and M.~Smolkin, {\it {Entanglement Entropy: A Perturbative
  Calculation}},  \href{http://xxx.lanl.gov/abs/1403.3733}{{\tt
  arXiv:1403.3733}}.

\bibitem{Lee:2014zaa}
J.~Lee, A.~Lewkowycz, E.~Perlmutter, and B.~R. Safdi, {\it {Renyi entropy,
  stationarity, and entanglement of the conformal scalar}},
  \href{http://xxx.lanl.gov/abs/1407.7816}{{\tt arXiv:1407.7816}}.

\bibitem{Klebanov:2012va}
I.~R. Klebanov, T.~Nishioka, S.~S. Pufu, and B.~R. Safdi, {\it {Is Renormalized
  Entanglement Entropy Stationary at RG Fixed Points?}},  {\em JHEP} {\bf 1210}
  (2012) 058, [\href{http://xxx.lanl.gov/abs/1207.3360}{{\tt
  arXiv:1207.3360}}].

\bibitem{Nishioka:2014kpa}
T.~Nishioka, {\it {Relevant Perturbation of Entanglement Entropy and
  Stationarity}},  {\em Phys.Rev.} {\bf D90} (2014) 045006,
  [\href{http://xxx.lanl.gov/abs/1405.3650}{{\tt arXiv:1405.3650}}].

\bibitem{Shiba:2013jja}
N.~Shiba and T.~Takayanagi, {\it {Volume Law for the Entanglement Entropy in
  Non-local QFTs}},  {\em JHEP} {\bf 1402} (2014) 033,
  [\href{http://xxx.lanl.gov/abs/1311.1643}{{\tt arXiv:1311.1643}}].

\bibitem{Pang:2014tpa}
D.-W. Pang, {\it {On holographic entanglement entropy of non-local field
  theories}},  {\em Phys.Rev.} {\bf D89} (2014) 126005,
  [\href{http://xxx.lanl.gov/abs/1404.5419}{{\tt arXiv:1404.5419}}].

\end{thebibliography}\endgroup

\end{document}